\documentclass{article}

\usepackage{arxiv}

\usepackage[utf8]{inputenc} % allow utf-8 input
\usepackage[T1]{fontenc}    % use 8-bit T1 fonts
\usepackage{hyperref}       % hyperlinks
\usepackage{url}            % simple URL typesetting
\usepackage{booktabs}       % professional-quality tables
\usepackage{amsfonts}       % blackboard math symbols
\usepackage{nicefrac}       % compact symbols for 1/2, etc.
\usepackage{microtype}      % microtypography
\usepackage{lipsum}
\usepackage{lscape}
\usepackage[authoryear, sort]{natbib}

\usepackage{morefloats}
\usepackage{color}
\usepackage{multirow}
\usepackage{latexsym}
\usepackage{amsmath,amssymb,amsthm,graphicx}
\usepackage{dsfont}
\usepackage{enumitem}

\newtheoremstyle{thm}% name
{9pt}%      Space above, empty = `usual value'
{9pt}%      Space below
{\itshape}% Body font
{}%         Indent amount (empty = no indent, \parindent = para indent)
{\bfseries}% Thm head font
{.}%        Punctuation after thm head
{ }% Space after thm head: \newline = linebreak
{}%         Thm head spec
\theoremstyle{thm}

\newtheoremstyle{def}% name
{9pt}%      Space above, empty = `usual value'
{9pt}%      Space below
{}% Body font
{}%         Indent amount (empty = no indent, \parindent = para indent)
{\bfseries}% Thm head font
{.}%        Punctuation after thm head
{ }% Space after thm head: \newline = linebreak
{}%         Thm head spec
\theoremstyle{def}

\renewcommand{\footnoterule}{%
	\kern -3.5pt
	\hrule width \textwidth height 1pt
	\kern 3.5pt
}

\makeatletter
\def\blfootnote{\xdef\@thefnmark{}\@footnotetext}
\makeatother

\title{Kaplan-Meier based tests for exponentiality in the presence of censoring}

\author{E. Bothma\\
Subject Group Statistics,\\ North-West University, South Africa. \\
\href{mailto:elzanieb@yahoo.com}{mailto:elzanieb@yahoo.com}\\
\And J.S. Allison\\
Subject Group Statistics,\\ North-West University, South Africa. \\
\href{mailto:james.allison@nwu.ac.za}{james.allison@nwu.ac.za}\\
\And M. Cockeran\\
Subject Group Statistics,\\ North-West University, South Africa. \\
\href{mailto:marike.cockeran@nwu.ac.za}{marike.cockeran@nwu.ac.za}\\
\And I.J.H. Visagie\\
Subject Group Statistics,\\ North-West University, South Africa. \\
\href{mailto:jaco.visagie@nwu.ac.za}{jaco.visagie@nwu.ac.za}\\
}

\begin{document}

\date{\today}
\maketitle

\begin{abstract}
In this paper we test the composite hypothesis that lifetimes follow an exponential distribution based on observed randomly right censored data. Testing this hypothesis is complicated by the presence of this censoring, due to the fact that not all lifetimes are observed. To account for this complication, we propose modifications to tests based on the empirical characteristic function and Laplace transform. In the full sample case these empirical functions can be expressed as integrals with respect to the empirical distribution function of the lifetimes. We propose replacing this estimate of the distribution function by the Kaplan-Meier estimate. The resulting test statistics can be expressed in easily calculable forms in terms of summations of functionals of the observed data. Additionally, a general framework for goodness-of-fit testing, in the presence of random right censoring, is outlined. A Monte Carlo study is performed, the results of which indicate that the newly modified tests generally outperform the existing tests. A practical application, concerning initial remission times of leukemia patients, is discussed along with some concluding remarks and avenues for future research.
\end{abstract}

\keywords{Exponential distribution,\ Goodness-of-fit testing,\ Hypothesis testing,\ Random right censoring,\ Warp-speed bootstrap.}

\section{Introduction}
\label{Intro}

The exponential distribution plays a central role in various fields such as survival analysis and reliability theory, see \cite{klein2006survival}. In the complete sample case, several tests for testing the hypothesis that the observed lifetimes are realisations from the exponential distribution have been developed. For an in-depth discussion of these tests, the reader is referred to \cite{allison2017apples} as well as \cite{henze2005recent} and the references therein. In the mentioned fields random right censoring often arises due to the nature of the study itself. Consider, for example, a medical study where the aim is to observe the lifetimes of patients with a specific disease. It often happens that a given patient is still alive at the end of the study or the patient leaves the study due to some other reason, such as dying in a car accident. Testing whether these observed lifetimes are realisations from a specified distribution is complicated by the fact that not all of these times are observed.

There is a relative scarcity of tests for exponentiality (or any other lifetime distribution) in the presence of random right censoring. One approach is to transform the censored sample to a complete sample and then use any of the existing tests for exponentiality developed for the full sample case. This approach is discussed in \cite{balakrishnan2015empirical}. A more common approach is to use the standard tests applicable to complete samples and modify them to accommodate random right censoring. \cite{koziol1976cramer} derived a modification of the Cram\'er-von Mises test statistic in the case of a simple hypothesis, which is based on the Kaplan-Meier product limit estimate of the distribution function. \cite{kim2012testing} studied modified versions of the Kolmogorov-Smirnov and Cram\'er-von Mises test statistics for testing a composite hypothesis. However, this study is limited by the fact that they are based on the random censoring model proposed by \cite{koziol1976cramer}. This implies that, under the null hypothesis, the censoring distribution is known, which is an unrealistic assumption in practise. It is further well known that the null distribution of the test statistics are dependent on the unknown censoring distribution (\citep[see, e.g.][]{d1986goodness}). Another test, specifically developed for the random censoring case, is the test proposed by \cite{cox1984analysis}, which is based on a score function. This test has also been shown to be quite powerful in the full sample case, see \cite{allison2017apples}.

A number of powerful tests for exponentiality are based on either the empirical characteristic function or the empirical Laplace transform. These include tests proposed in \cite{epps1986test}, \cite{baringhaus1991class}, \cite{henze2002goodness} and \cite{henze2002tests}. For an overview of testing procedures involving the characteristic function see the discussion paper by \cite{meintanis2016review}. In this paper we propose newly modified versions of these tests that can be used in the presence of random right censoring. We assume that the censoring distribution is unknown and that it is estimated non-parametrically.

Before proceeding some notation is introduced. Let $X_1, \dots, X_n$ be independent and identically distributed (i.i.d.) lifetime variables with continuous distribution function $F$  and let $C_1, \dots, C_n$ be i.i.d. censoring variables with distribution function $G$, independent of $X_1, \dots, X_n$. Let
$$T_j=\mbox{min}(X_j,C_j) \ \mbox{and} \ \delta_j=
    \begin{cases}
      1, & \text{if}\ X_j\leq C_j \\
      0, & \text{if}\ X_j > C_j.
    \end{cases}$$
Based on the observed pairs $(T_j, \delta_j), \ j=1, \dots n$ we wish to test the composite hypothesis 
\begin{equation}\label{hypothesis}
H_0: F \text{ is the exponential distribution with expectation } 1/\lambda,
\end{equation}
for some unknown $\lambda>0$ against non-exponential alternatives. Denote the order statistics of $X_1, \dots ,X_n$ and  $T_1, \dots ,T_n$  by $X_{(1)} < X_{(2)} < \cdots < X_{(n)}$ and $T_{(1)} < T_{(2)} < \cdots < T_{(n)}$, respectively. Note that $\delta_{(j)}$ represents the indicator variable corresponding to $T_{(j)}$. Using the notation introduced above, the Kaplan-Meier estimator, $\tilde{F}_n$, of the distribution function, is given by

\begin{equation}
    1-\tilde{F}_n(t)=
    \begin{cases}
      1, &  t \leq T_{(1)} \\
      \prod_{j=1}^{k-1}\left(\frac{n-j}{n-j+1}\right)^{\delta_{(j)}}, & T_{(k-1)}< t \leq T_{(k)},  \ \ \ \ k=2,\dots, n. \nonumber\\
      \prod_{j=1}^n \left(\frac{n-j}{n-j+1}\right)^{\delta(j)}, & t>T_{(n)}.
    \end{cases}
  \end{equation}
For more details about this estimator see \cite{kaplan1958nonparametric} and \cite{efron1967two}.
%, and \cite{breslow1974large}.

All of the test statistics under consideration make use of scaled lifetime values, denoted by $Y_j = T_j\hat{\lambda}$, where $\hat{\lambda} = \sum_{j=1}^n\delta_j/\sum_{j=1}^n T_j$ is the maximum likelihood estimate of $\lambda$. The invariance property of the exponential distribution justifies the use of these scaled values \citep[see, e.g.][]{gupta1997invariance}.

The remainder of the paper is structured as follows. In Section \ref{Tests} we indicate how four existing tests are modified to accommodate random right censoring. Since the null distribution of each of the test statistics depends on the unknown censoring distribution, we propose a parametric bootstrap procedure in Section \ref{Bootstrap} in order to compute critical values for the tests under consideration. Section \ref{MonteCarlo} contains the results of a Monte Carlo study where the empirical powers of the newly modified tests are compared to those of existing tests. The paper concludes in Section \ref{Conclude} with an application to observed leukemia remission times as well as some avenues for future research.

\section{Proposed test statistics}
\label{Tests}

The statistical literature contains several tests for exponentiality based on the empirical characteristic function and Laplace transform. The tests below are modifications of these tests obtained by estimating the mentioned functions using the Kaplan-Meier estimate of the distribution function instead of the empirical distribution function. Below we consider a test based on the empirical characteristic function before turning our attention to tests based on the empirical Laplace transform.

Recall that the characteristic function of a random variable, $X$, with distribution, $F$, is defined to be
$$\phi(t) = E\left[\textrm{e}^{itY}\right] = \int \textrm{e}^{ity} \mathrm{d}F(y),$$
where $i=\sqrt{-1}$. Denote the characteristic function of the standard exponential distribution by $\phi(t) = (1+t)^{-1}$. Let $F_n$ denote the empirical distribution function of the random variables $Y_1, \dots, Y_n$
\begin{equation*}
    F_n(y) = \sum_{j=1}^n \textrm{I}(Y_j \leq y),
\end{equation*}
where $\textrm{I}$ denotes the indicator function. Using $F_n$
to estimate $F$, we obtain the empirical characteristic function 
$$\phi_n(t) = \int \textrm{e}^{ity}\mathrm{d}F_n(y) = \frac{1}{n}\sum_{j=1}^n \textrm{e}^{itY_j}.$$
Upon replacing $F_n$ by $\tilde{F}_n$, we obtain an estimate for the characteristic function based on a censored sample
$$\tilde{\phi}_n(t) = \int \textrm{e}^{ity}\mathrm{d}\tilde{F}_n(y) = \sum_{j=1}^n \Delta_j \textrm{e}^{itY_j},$$
where $\Delta_j$ denotes the size of the jump in $\tilde{F}_n(T_{(j)}), j=1,\dots,n$, given by
$$\Delta_1 = \frac{\delta_{(1)}}{n}, \text{ } \Delta_n = \prod_{j=1}^{n-1}\left(\frac{n-j}{n-j+1}\right)^{\delta_{(j)}} \text{ and}$$
\begin{align*}
\Delta_j &= \prod_{k=1}^{j-1} \left(\frac{n-k}{n-k+1}\right)^{\delta_{(k)}} - \prod_{k=1}^{j} \left(\frac{n-k}{n-k+1}\right)^{\delta_{(k)}} = \frac{\delta_{(j)}}{n-j+1} \prod_{k=1}^{j-1} \left(\frac{n-k}{n-k+1}\right)^{\delta_{(k)}}, \ j=2,\dots,n-1.
\end{align*}

\cite{epps1986test} introduced a test statistic based on the difference between the characteristic function and its empirical counterpart. The proposed test statistic is given by
\begin{equation*}
 EP_n = \frac{1}{2\pi}\int \left[\phi_n(t)-\phi(t)\right]\phi(-t) \mathrm{d}t.
\end{equation*}
In the presence of random right censoring, we modify the test statistic by replacing $\phi_n$ with $\tilde{\phi_n}$. The resulting modified test statistic is given by
\begin{equation*}
 \widetilde{EP}_n = \frac{1}{2\pi} \int \left[\tilde{\phi}_n(t)-\phi(t)\right]\phi(-t) \mathrm{d}t.
\end{equation*}
After straightforward calculations, this test statistic can be expressed as
\begin{equation*}
 \widetilde{EP}_n = \sqrt{48n}\left[\sum_{j=1}^n \Delta_j \textrm{e}^{-Y_j} - \frac{1}{2}\right].
\end{equation*}
The null hypothesis in \eqref{hypothesis} is rejected for large values of $|\widetilde{EP}_n|$.

We now turn our attention to tests based on the empirical Laplace transform. Using similar notation to that used for the various versions of the characteristic function, let $\psi$, $\psi_n$ and $\tilde{\psi}_n$ respectively denote the Laplace transform, the empirical Laplace transform and the empirical version of this function obtained using the Kaplan-Meier estimate of the distribution. The resulting functions can be expressed as
\begin{equation*}
    \psi(t) = \int \textrm{e}^{-ty} \mathrm{d}F(y), \ \ \ \ \ \psi_n(t) = \frac{1}{n}\sum_{j=1}^n \textrm{e}^{-tY_j} \ \ \ \ \ \text{and} \ \ \ \ \ \tilde{\psi}_n(t) = \sum_{j=1}^n \Delta_j \textrm{e}^{-tY_j}.
\end{equation*}

\cite{baringhaus1991class} proposed a test based on a partial differential equation involving the Laplace transform. In the presence of random right censoring, the modified version of this test becomes
\begin{equation*}
 \tilde{B}_{n,a} = n\int_{0}^{\infty} \left[(1+t)\tilde{\psi}_n^{'}(t) + \tilde{\psi}_n(t)\right]^2 \textrm{e}^{-at} \mathrm{d}t, 
\end{equation*}
where $a>0$ is a user-specified tuning parameter. After some algebra, this test statistic can be expressed as
\begin{equation*}
  \tilde{B}_{n,a} = n \sum_{j=1}^n \sum_{k=1}^n \Delta_j\Delta_k \left[\frac{(1-Y_j)(1-Y_k)}{Y_j+Y_k+a} - \frac{Y_j+Y_k}{(Y_j+Y_k+a)^2} + \frac{2Y_jY_k}{(Y_j+Y_k+a)^2} + \frac{2Y_jY_k}{(Y_j+Y_k+a)^3}\right].
\end{equation*}
The null hypothesis of exponentiality is rejected for large values of $\tilde{B}_{n,a}$.

Another test involving the Laplace transform was proposed in \cite{henze2002tests}. This test statistic is based on the squared difference between the Laplace transform and its empirical counterpart. In the presence of random censoring this test can be modified to have test statistic
\begin{equation*}
 \tilde{L}_{n,a} = n\int_{0}^{\infty} \left[\tilde{\psi}_n(t)-\psi(t)\right]^2 (1+t)^2\textrm{e}^{-at} \mathrm{d}t, 
\end{equation*}
with the following form that can easily be implemented
\begin{equation*}
    \tilde{L}_{n,a} = n \sum_{j=1}^n \sum_{k=1}^n \Delta_j\Delta_k \left[\frac{1+(Y_j+Y_k+a+1)^2}{(Y_j+Y_k+a)^3}\right] - 2n \sum_{j=1}^n \Delta_j \left[\frac{1+Y_j+a}{(Y_j+a)^2}\right] + \frac{n}{a}.
\end{equation*}
The null hypothesis is rejected for large values of $\tilde{L}_{n,a}$.

\cite{henze2002goodness} proposed a goodness-of-fit test based on a characterisation of the exponential distribution via the characteristic function. Modifying the proposed test, we obtain the test statistic
\begin{equation*}
 \tilde{H}_{n,a} = n\int_{0}^{\infty} \left[S_n(t)-tC_n(t)\right]^2 \textrm{e}^{-at} \mathrm{d}t, 
\end{equation*}
where $S_n(t)=\sum_{j=1}^n \Delta_j \sin(tY_j)$ and $C_n(t)=\sum_{j=1}^n \Delta_j \cos(tY_j)$.
This statistic admits the following easily calculable form
\begin{align*}
 \tilde{H}_{n,a} &= \frac{an}{2} \sum_{j=1}^n \sum_{k=1}^n \Delta_j\Delta_k \left[\frac{1}{a^2+\left(Y_j-Y_k\right)^2} - \frac{1}{a^2+\left(Y_j+Y_k\right)^2} - \frac{4\left(Y_j+Y_k\right)}{\left(a^2+\left(Y_j+Y_k\right)^2\right)^2}\right.\\
 &+ \left. \frac{2a^2-6\left(Y_j-Y_k\right)^2}{\left(a^2+\left(Y_j-Y_k\right)^2\right)^3} + \frac{2a^2-6\left(Y_j+Y_k\right)^2}{\left(a^2+\left(Y_j+Y_k\right)^2\right)^3}\right].
\end{align*}
The null hypothesis is rejected for large values of $\tilde{H}_{n,a}$.

\section{Bootstrap algorithm}
\label{Bootstrap}
The null distribution of each of the test statistics considered depends on the unknown censoring distribution, even in the case of a simple hypothesis \citep[see][]{d1986goodness}. Since we will not assume any known form of the censoring distribution (e.g. the Kozoil-Green model), we propose the following parametric bootstrap algorithm to estimate the critical values of the tests.
\begin{enumerate}
	\item Based on the pairs $(T_j,\delta_j), j=1,\dots,n$ estimate $\lambda$ by $\hat{\lambda} = \sum\delta_j/\sum T_j$. 
	\item Obtain a parametric bootstrap sample $X_1^*, X_2^*, \dots, X_n^*$ by sampling from an exponential distribution with parameter $\hat{\lambda}$. 
	\item Obtain a non-parametric bootstrap sample $C_1^*, C_2^*, \dots, C_n^*$ by sampling from the Kaplan-Meier estimate of the distribution of the censoring times. 
	\item Set $$T_j^*=\mbox{min}(X_j^*,C_j^*) \ \mbox{and} \ \delta_j^*=
    \begin{cases}
      1, & \text{if}\ X_j^*\leq C_j^* \\
      0, & \text{if}\ X_j^* > C_j^*.
    \end{cases}$$ 
	\item Calculate $\hat{\lambda}^* = \sum\delta_j^*/\sum T_j^*$ and obtain the scaled bootstrap values $Y_j^*=T_j^*\hat{\lambda}^*$.
	\item Based on the data $\left(Y_j^*,\delta_j^*\right), \ j=1,\dots,n$ calculate the value of the test statistic, say\\$S^* = S\left(\left(Y_1^*,\delta_1^*\right), \left(Y_2^*,\delta_2^*\right), \dots, \left(Y_n^*,\delta_n^*\right)\right)$. 
	\item Repeat steps 2-6 B times to obtain $S_1^*, S_2^*, \dots, S_B^*$. Obtain the order statistics, $S_{(1)}^* \leq S_{(2)}^* \leq \dots \leq S_{(B)}^*$. The estimated critical value is then $\hat{c}_n(\alpha) = S_{\lfloor B(1-\alpha)\rfloor}^*$ where $\lfloor A\rfloor$ denotes the floor of $A$.
\end{enumerate}
The algorithm provided above is quite general and can easily be amended in order to test for any lifetime distribution in the presence of random censoring.

\section{Monte Carlo study}
\label{MonteCarlo}

In this section the power behaviour of the newly modified tests is investigated by means of a Monte Carlo study. The  $\widetilde{EP}_n$, $\widetilde{L}_{n,a}$, $\widetilde{B}_{n,a}$ and $\widetilde{H}_{n,a}$ tests are compared to the modified Kolmogorov-Smirnov ($\widetilde{KS}_n$) and Cram\'er-von Mises  ($\widetilde{CM}_n$) tests proposed in \cite{koziol1976cramer} as well as the \cite{cox1984analysis} test ($\widetilde{CO}_n$), which was originally proposed for use with censored data. 

Let $\tilde{n}$ denote the number of observations which are uncensored, i.e., $\tilde{n}=\sum \delta_j$. The calculable forms of the Kolmogorov-Smirnov, Cram\'er-von Mises and Cox and Oakes tests are, respectively, given by
\begin{eqnarray*}
\widetilde{KS}_n &=& \sup_{x \geq 0} \left|\tilde{F}_n(x)-\left(1-\textrm{e}^{-x}\right)\right| \\
 &=& \max\left[\max_{1\leq j \leq n} \left\{\tilde{F}_n(Y_{(j)})-\left(1-\textrm{e}^{-Y_{(j)}}\right)\right\},\max_{1\leq j \leq n} \left\{\left(1-\textrm{e}^{-Y_{(j)}}\right)-\tilde{F}_n^-(Y_{(j)})\right\}\right],
\end{eqnarray*}

\begin{align*}
\widetilde{CM}_n &= n\int_0^1 \left(t-\tilde{F}_n(t)\right)^2 \mathrm{d}t = \frac{n}{3} + n\sum_{j=1}^{\tilde{n}+1} \tilde{F}_n\left(\tilde{Y}_{j-1}\right) \left(\tilde{Y}_j-\tilde{Y}_{j-1}\right) \left[\tilde{F}_n\left(\tilde{Y}_{j-1}\right)-\left(\tilde{Y}_j+\tilde{Y}_{j-1}\right)\right] \ \ \text{ and}
\end{align*}

\begin{align*}
\widetilde{CO}_n &= \tilde{n} + \sum_{j=1}^n \log(Y_j)\delta_j - \frac{\tilde{n} \sum_{j=1}^n Y_j\log(Y_j)}{\sum_{j=1}^n Y_j}.
\end{align*}

The null hypothesis of exponentiality is rejected in the case of large values of $\widetilde{KS}_n$ and $\widetilde{CM}_n$, while the Cox and Oakes test rejects for small or large values of $\widetilde{CO}_n$.

\subsection{Simulation setting}
The nominal significance level is set to 5\% throughout. Empirical powers are presented for sample sizes of $n=50$ and $n=100$. The reported empirical powers are calculated in the case of 10\%, 20\% and 30\% censoring. For each lifetime distribution considered, we include the power achieved using three different censoring distributions; the exponential, uniform and Lindley distributions. The alternative distributions used are listed in Table \ref{table1}.

A warp-speed bootstrap methodology \citep{GPW:2013}, which essentially entails using a single bootstrap replication for each Monte Carlo sample generated, is employed in order to calculate empirical powers. This methodology is used in order to reduce the computational cost of the calculation of these powers and has been employed by a number of authors in the literature to compare Monte Carlo performances \citep[see, e.g.][]{HMP2020}.

The estimated powers are reported in Tables \ref{table2} to \ref{table7}. In the interest of continuity, these tables are deferred to the end of the paper. The reported empirical powers represent the percentages of 50\,000 independent Monte Carlo samples that resulted in the rejection of the null hypothesis (rounded to the nearest integer). For each test considered Tables \ref{table2} to \ref{table7} show three empirical powers against each of the lifetime distributions, corresponding to the three different censoring distributions used. In each case, the results for the exponential, uniform and Lindley distribution are shown in the first, second and third lines, respectively. For the reader’s comfort the highest power against each alternative distribution is displayed in bold in the simulation results.

In the discussion below, including the tables, the subscript $n$ is suppressed. For $\widetilde{L}$ and $\widetilde{B}$, we include numerical powers in the cases where $a$ is set to $0.25$ and $0.5$, while $a=0.5$ and $a=1$ are used in conjunction with $\widetilde{H}$. All calculations were performed in \textsf{R} \citep{CRAN}. The $\emph{LindleyR}$ package was used to generate samples from censored distributions, see \cite{Lind}. The tables were produced using the $\emph{Stargazer}$ package \cite{Starg}.

\begin{table}[!htbp]
	\centering\footnotesize
	\caption{Density functions of the alternative distributions.\label{table1}}
	\begin{tabular}{|c|c|c|}
	\hline
	Alternative & Density & Notation \\
	\hline
	Exponential & $\theta \exp(-\theta x)$ & $Exp(\theta)$  \\
	Gamma & $\left(\Gamma(\theta)\right)^{-1}x^{\theta-1}\exp(-x)$ & $\Gamma(\theta)$\\
	Weibull & $\theta x^{\theta-1}\exp(-x^{\theta})$ & $W(\theta)$  \\
	Lognormal & $\left(\theta x \sqrt{2\pi}\right)^{-1} \exp\left(-{\log^2(x)}\left(2\theta^2\right)^{-1}\right)$ & $LN(\theta)$\\
	Chi square & $\left(2^{\theta/2}\Gamma(\theta/2)\right)^{-1}x^{\theta/2-1}\exp(-x/2)$ & $\chi^2(\theta)$  \\
	Beta & ${x^{\alpha-1}(1-x)^{\theta-1}\Gamma(\alpha+\theta)}\left(\Gamma(\alpha)\Gamma(\theta)\right)^{-1}$ & $\beta(\alpha,\theta)$\\	
	\hline
	\end{tabular}\normalsize
\end{table} 

\subsection{Simulation results}

The size of the tests are maintained closely for all sample sizes and for censoring proportions of 10\% and 20\%, perhaps with the single exception of $\widetilde{B}$ in the case of 20\% censoring. For 30\% censoring we find that $\widetilde{KS}$, $\widetilde{CM}$, $\widetilde{EP}$ and $\widetilde{B}$ are all slightly conservative, obtaining estimated sizes ranging between 2\% and 4\%. For small censoring proportions, the effect of the censoring distribution on the power of a given test is not pronounced. However, as the censoring proportion increases, these differences become more noticeable. As one would expect, the powers generally increase with the sample size and decrease slightly as the censoring proportion increases. We also estimated the powers for the complete sample case, the results that we obtained are similar to those of \cite{allison2017apples} and are not included below. The choice of the tuning parameter, $a$, also plays a role in the performance of $\widetilde{L}$, $\widetilde{B}$ and $\widetilde{H}$. Overall, these tests have higher powers for smaller values of $a$.

In general, the $\widetilde{CO}$ and $\widetilde{L}$ tests outperform their competitors. As a result, we recommend using either the $\widetilde{CO}$ or $\widetilde{L}_{.25}$ test in practise. It is interesting to note that, when we disregard the $\widetilde{CO}$, the newly modified tests clearly outperform the modified tests based on the distribution function.

\section{Practical application and conclusion}
\label{Conclude}
In this section, we apply each of the tests discussed in Section \ref{MonteCarlo} to a real-world data set. The initial remission times of leukemia patients, in days, were reported in \cite{lee2003statistical}, this data set can be found in Table \ref{table8}. An asterisk is used in order to indicate that the observation is censored. The original data were segmented into three treatment groups. However, \cite{lee2003statistical} showed that the data do not display significant differences among the various treatments. As a result we treat the data as i.i.d. realisations from a single, censored, lifetime distribution. The estimated p-values for each of the tests, obtained using 1 million bootstrap replications, are displayed in Table \ref{table8}.

\begin{table}[!htbp] \centering 
  \caption{Initial remission times of leukemia patients, in days.} 
  \label{table8} 
\begin{tabular}{@{\extracolsep{5pt}} c} 
\hline
$4,5,8,8,9,10,10,10,10,10,11,12,12,12^*,13,14,20,20^*,23,23,25,25,25,28,28,28,$\\
$28,29,31,31,31,32,37,40,41,41,48,48,57,62,70,74,75,89,99,100,103,124,139,143,$\\
$159^*,161^*,162,169,190^*,195,196^*,197^*,199^*,205^*,217^*,219^*,220,245^*,258^*,269^*$\\
\hline
\end{tabular} 
\end{table}

\begin{table}[!htbp] \centering 
  \caption{$p$-values associated with the various tests used.} 
  \label{table9} 
\begin{tabular}{@{\extracolsep{1pt}} ccccccccccc} 
\\[-1.8ex]\hline \vspace{-0.3cm}\\
$Test$ & $\widetilde{KS}$ & $\widetilde{CM}$ & $\widetilde{CO}$ & $\widetilde{EP}$ & $\widetilde{L}_{.25}$ & $\widetilde{L}_{.5}$ & $\widetilde{B}_{.25}$ & $\widetilde{B}_{.5}$ & $\widetilde{H}_{.5}$ & $\widetilde{H}_{1}$ \\
\hline 
$p-value$ & $<0.01$ & $<0.01$ & $0.03$ & $0.11$ & $0.13$ & $0.03$ & $<0.01$ & $<0.01$ & $0.06$ & $<0.01$ \\ 
\hline 
\end{tabular}
\end{table}

The results from the practical example show that each of the tests, except $\widetilde{EP}$, $\widetilde{L}_{.25}$ and $\widetilde{H}_{.5}$, reject the null hypothesis of exponentiality of the initial remission times at a 5\% significance level. This provides some evidence that a more flexible distribution, perhaps with heavier tails, should be used as a model for this specific data set.

In this paper we modify four tests for exponentiality, based on either the characteristic function or the Laplace transform, to accommodate random right censoring. These tests are compared to existing tests using a Monte Carlo study and it is found that the newly modified tests are competitive in terms of their power performance. This is especially true for the test based on the squared difference between the Laplace transform and its empirical counterpart, $\widetilde{L}$, which, together with the $\widetilde{CO}$ test, is recommended for use in practise.

Although we only consider these tests in the i.i.d. setup, these modified tests can also be used to test the adequacy of fit of various survival models when random right censoring is present. These models include the conventional Cox proportional hazards model as well as the mixture cure model, which is currently receiving a lot of attention in the literature, see \cite{amico2018cure}. Recently, \cite{betsch2019new} introduced a novel idea of basing goodness-of-fit tests on a fixed point property of a new transformation connected to the Stein characterization. These tests were found to be quite powerful in the full sample case and it will be interesting to see whether they can be successfully modified to the censored case.

%\backmatter

\section*{Acknowledgments}

The financial assistance of the National Research Foundation (NRF) towards
this research is hereby acknowledged. Opinions expressed and conclusions
arrived at, are those of the authors and are not necessarily to be
attributed to the NRF.

%\nocite{*}% Show all bib entries - both cited and uncited; comment this line to view only cited bib entries;
\bibliography{Article1}%

\begin{thebibliography}{}

\bibitem[Allison et~al., 2017]{allison2017apples}
Allison, J.~S., Santana, L., Smit, N., and Visagie, I. J.~H. (2017).
\newblock An "apples-to-apples" comparison of various tests for exponentiality.
\newblock {\em Computational Statistics}, 32(4):1241--1283.

\bibitem[Amico and Van~Keilegom, 2018]{amico2018cure}
Amico, M. and Van~Keilegom, I. (2018).
\newblock Cure models in survival analysis.
\newblock {\em Annual Review of Statistics and Its Application}, 5:311--342.

\bibitem[Balakrishnan et~al., 2015]{balakrishnan2015empirical}
Balakrishnan, N., Chimitova, E., and Vedernikova, M. (2015).
\newblock An empirical analysis of some nonparametric goodness-of-fit tests for
  censored data.
\newblock {\em Communications in Statistics-Simulation and Computation},
  44(4):1101--1115.

\bibitem[Baringhaus and Henze, 1991]{baringhaus1991class}
Baringhaus, L. and Henze, N. (1991).
\newblock A class of consistent tests for exponentiality based on the empirical
  {L}aplace transform.
\newblock {\em Annals of the Institute of Statistical Mathematics},
  43(3):551--564.

\bibitem[Betsch and Ebner, 2019]{betsch2019new}
Betsch, S. and Ebner, B. (2019).
\newblock A new characterization of the gamma distribution and associated
  goodness-of-fit tests.
\newblock {\em Metrika}, 82(7):779--806.

\bibitem[Cox and Oakes, 1984]{cox1984analysis}
Cox, D.~R. and Oakes, D. (1984).
\newblock {\em Analysis of {S}urvival {D}ata}, volume~21.
\newblock CRC Press.

\bibitem[D'Agostino and Stephens, 1986]{d1986goodness}
D'Agostino, R.~B. and Stephens, M.~A. (1986).
\newblock {\em Goodness-of-fit {T}echniques}, volume~68.
\newblock CRC press.

\bibitem[Efron, 1967]{efron1967two}
Efron, B. (1967).
\newblock The two sample problem with censored data.
\newblock In {\em Proceedings of the Fifth Berkeley Symposium on Mathematical
  Statistics and Probability}, volume~4, pages 831--853.

\bibitem[Epps and Pulley, 1986]{epps1986test}
Epps, T.~W. and Pulley, L.~B. (1986).
\newblock A test of exponentiality vs. monotone-hazard alternatives derived
  from the empirical characteristic function.
\newblock {\em Journal of the Royal Statistical Society: Series B
  (Methodological)}, 48(2):206--213.

\bibitem[Giacomini et~al., 2013]{GPW:2013}
Giacomini, R., Politis, D.~N., and White, H. (2013).
\newblock A warp-speed method for conducting {M}onte {C}arlo experiments
  involving bootstrap estimators.
\newblock {\em Econometric Theory}, 29(3):567--589.

\bibitem[Gupta and Richards, 1997]{gupta1997invariance}
Gupta, R.~D. and Richards, D. S.~P. (1997).
\newblock Invariance properties of some classical tests for exponentiality.
\newblock {\em Journal of statistical planning and inference}, 63(2):203--213.

\bibitem[Henze and Meintanis, 2002a]{henze2002goodness}
Henze, N. and Meintanis, S.~G. (2002a).
\newblock Goodness-of-fit tests based on a new characterization of the
  exponential distribution.
\newblock {\em Communications in Statistics-Theory and Methods},
  31(9):1479--1497.

\bibitem[Henze and Meintanis, 2002b]{henze2002tests}
Henze, N. and Meintanis, S.~G. (2002b).
\newblock Tests of fit for exponentiality based on the empirical {L}aplace
  transform.
\newblock {\em Statistics}, 36(2):147--161.

\bibitem[Henze and Meintanis, 2005]{henze2005recent}
Henze, N. and Meintanis, S.~G. (2005).
\newblock Recent and classical tests for exponentiality: a partial review with
  comparisons.
\newblock {\em Metrika}, 61(1):29--45.

\bibitem[Hlavac, 2018]{Starg}
Hlavac, M. (2018).
\newblock {\em stargazer: Well-formatted regression and summary statistics
  tables}.

\bibitem[Hu{\v{s}}kov{\'a} et~al., 2020]{HMP2020}
Hu{\v{s}}kov{\'a}, M., Meintanis, S.~G., and Pretorius, C. (2020).
\newblock Tests for validity of the semiparametric heteroskedastic
  transformation model.
\newblock {\em Computational Statistics \& Data Analysis}, 144.

\bibitem[Kaplan and Meier, 1958]{kaplan1958nonparametric}
Kaplan, E.~L. and Meier, P. (1958).
\newblock Nonparametric estimation from incomplete observations.
\newblock {\em Journal of the American Statistical Association},
  53(282):457--481.

\bibitem[Kim, 2012]{kim2012testing}
Kim, N.-H. (2012).
\newblock Testing exponentiality based on {EDF} statistics for randomly
  censored data when the scale parameter is unknown.
\newblock {\em The Korean Journal of Applied Statistics}, 25(2):311--319.

\bibitem[Klein and Moeschberger, 2006]{klein2006survival}
Klein, J.~P. and Moeschberger, M.~L. (2006).
\newblock {\em Survival Analysis: {T}echniques for {C}ensored and {T}runcated
  {D}ata}.
\newblock Springer Science \& Business Media.

\bibitem[Koziol and Green, 1976]{koziol1976cramer}
Koziol, J.~A. and Green, S.~B. (1976).
\newblock A {C}ram{\'e}r-von {M}ises statistic for randomly censored data.
\newblock {\em Biometrika}, 63(3):465--474.

\bibitem[Lee and Wang, 2003]{lee2003statistical}
Lee, E.~T. and Wang, J. (2003).
\newblock {\em Statistical {M}ethods for {S}urvival {D}ata {A}nalysis}, volume
  476.
\newblock John Wiley \& Sons.

\bibitem[Mazucheli et~al., 2016]{Lind}
Mazucheli, J., Fernandes, L.~B., and {de Oliveira}, R.~P. (2016).
\newblock {\em Lindley{R}: The {L}indley Distribution and Its Modifications}.
\newblock R package version 1.1.0.

\bibitem[Meintanis, 2016]{meintanis2016review}
Meintanis, S.~G. (2016).
\newblock A review of testing procedures based on the empirical characteristic
  function.
\newblock {\em South African Statistical Journal}, 50(1):1--14.

\bibitem[{R Core Team}, 2019]{CRAN}
{R Core Team} (2019).
\newblock {\em R: {A} Language and Environment for Statistical Computing}.
\newblock R Foundation for Statistical Computing, Vienna, Austria.

\end{thebibliography}
\bibliographystyle{apalike}

\begin{table}[!htbp] \centering 
  \caption{Estimated powers for $10\%$ censoring for a sample size of n=50 with three different censoring distributions.} 
  \label{table2} 
  \scriptsize
\begin{tabular}{@{\extracolsep{1pt}} ccccccccccc} 
\\[-1.8ex]\hline \vspace{-0.15cm}\\
$F$ & $\widetilde{KS}$ & $\widetilde{CM}$ & $\widetilde{CO}$ & $\widetilde{EP}$ & $\widetilde{L}_{.25}$ & $\widetilde{L}_{.5}$ & $\widetilde{B}_{.25}$ & $\widetilde{B}_{.5}$ & $\widetilde{H}_{.5}$ & $\widetilde{H}_{1}$ \\
\hline 
\multirow{3}[2]{*}{$Exp(1)$} & $5$ & $5$ & $5$ & $5$ & $5$ & $5$ & $5$ & $4$ & $5$ & $5$ \\ 
 & $5$ & $5$ & $5$ & $5$ & $5$ & $5$ & $4$ & $4$ & $5$ & $5$ \\ 
 & $5$ & $5$ & $5$ & $5$ & $5$ & $5$ & $4$ & $4$ & $5$ & $5$ \\
\hline
\multirow{3}[2]{*}{$\Gamma(0.6)$} & $56$ & $63$ & $81$ & $63$ & $\mathbf{82}$ & $79$ & $80$ & $75$ & $55$ & $56$ \\ 
& $55$ & $62$ & $\mathbf{82}$ & $63$ & $\mathbf{82}$ & $80$ & $79$ & $75$ & $56$ & $55$ \\ 
& $55$ & $62$ & $\mathbf{82}$ & $63$ & $\mathbf{82}$ & $79$ & $79$ & $75$ & $56$ & $56$ \\ 
\hline
\multirow{3}[2]{*}{$\Gamma(0.8)$} & $14$ & $16$ & $25$ & $18$ & $\mathbf{27}$ & $25$ & $24$ & $22$ & $14$ & $14$ \\ 
& $14$ & $16$ & $25$ & $18$ & $\mathbf{27}$ & $24$ & $24$ & $22$ & $13$ & $14$ \\ 
& $14$ & $16$ & $25$ & $18$ & $\mathbf{28}$ & $25$ & $24$ & $21$ & $14$ & $15$ \\ 
\hline
\multirow{3}[2]{*}{$\Gamma(1.2)$} & $10$ & $12$ & $\mathbf{14}$ & $11$ & $11$ & $\mathbf{14}$ & $11$ & $10$ & $9$ & $11$ \\ 
 & $11$ & $12$ & $\mathbf{14}$ & $12$ & $12$ & $\mathbf{14}$ & $11$ & $10$ & $9$ & $11$ \\ 
 & $10$ & $11$ & $\mathbf{14}$ & $12$ & $11$ & $13$ & $10$ & $8$ & $9$ & $11$ \\ 
\hline
\multirow{3}[2]{*}{$W(0.8)$} & $32$ & $37$ & $\mathbf{51}$ & $42$ & $49$ & $48$ & $50$ & $47$ & $27$ & $31$ \\ 
 & $31$ & $37$ & $\mathbf{50}$ & $42$ & $49$ & $49$ & $48$ & $47$ & $27$ & $31$ \\ 
 & $31$ & $38$ & $\mathbf{51}$ & $41$ & $48$ & $48$ & $48$ & $46$ & $27$ & $31$ \\ 
\hline
\multirow{3}[2]{*}{$W(1.2)$} & $21$ & $25$ & $\mathbf{29}$ & $26$ & $23$ & $28$ & $23$ & $23$ & $16$ & $22$ \\ 
 & $20$ & $25$ & $\mathbf{30}$ & $26$ & $23$ & $27$ & $22$ & $22$ & $16$ & $23$ \\ 
 & $21$ & $25$ & $\mathbf{29}$ & $26$ & $22$ & $27$ & $21$ & $20$ & $15$ & $22$ \\ 
\hline
\multirow{3}[2]{*}{$LN(1)$} & $21$ & $26$ & $14$ & $13$ & $27$ & $20$ & $24$ & $18$ & $\mathbf{29}$ & $16$ \\ 
 & $19$ & $23$ & $12$ & $11$ & $28$ & $20$ & $24$ & $18$ & $\mathbf{28}$ & $15$ \\ 
 & $19$ & $23$ & $13$ & $11$ & $28$ & $20$ & $23$ & $16$ & $\mathbf{29}$ & $15$ \\ 
\hline
\multirow{3}[2]{*}{$LN(1.5)$} & $82$ & $87$ & $83$ & $\mathbf{90}$ & $70$ & $80$ & $85$ & $87$ & $59$ & $78$ \\ 
 & $81$ & $86$ & $82$ & $\mathbf{89}$ & $68$ & $79$ & $84$ & $87$ & $56$ & $76$ \\ 
 & $79$ & $84$ & $81$ & $\mathbf{88}$ & $66$ & $78$ & $83$ & $85$ & $54$ & $75$ \\ 
\hline
\multirow{3}[2]{*}{$\chi^2(1)$} & $82$ & $87$ & $\mathbf{97}$ & $86$ & $\mathbf{97}$ & $96$ & $96$ & $94$ & $83$ & $81$ \\ 
 & $81$ & $86$ & $\mathbf{97}$ & $86$ & $\mathbf{97}$ & $95$ & $96$ & $94$ & $82$ & $81$ \\ 
 & $82$ & $87$ & $\mathbf{97}$ & $86$ & $\mathbf{97}$ & $95$ & $95$ & $94$ & $83$ & $82$ \\ 
\hline
\multirow{3}[2]{*}{$\chi^2(3)$} & $33$ & $40$ & $\mathbf{51}$ & $40$ & $46$ & $50$ & $44$ & $42$ & $30$ & $37$ \\ 
 & $34$ & $40$ & $\mathbf{51}$ & $40$ & $45$ & $49$ & $42$ & $39$ & $29$ & $37$ \\ 
 & $34$ & $41$ & $\mathbf{50}$ & $39$ & $46$ & $\mathbf{50}$ & $39$ & $34$ & $29$ & $36$ \\ 
\hline
\multirow{3}[2]{*}{$\beta(1,1)$} & $89$ & $\mathbf{97}$ & $85$ & $95$ & $54$ & $75$ & $82$ & $88$ & $78$ & $\mathbf{97}$ \\ 
 & $89$ & $\mathbf{97}$ & $86$ & $95$ & $55$ & $74$ & $81$ & $87$ & $78$ & $\mathbf{97}$ \\ 
 & $89$ & $\mathbf{97}$ & $85$ & $95$ & $54$ & $74$ & $81$ & $87$ & $78$ & $\mathbf{97}$ \\ 
\hline
\multirow{3}[2]{*}{$\beta(0.5,1)$} & $35$ & $47$ & $49$ & $8$ & $\mathbf{69}$ & $54$ & $60$ & $43$ & $55$ & $49$ \\ 
 & $34$ & $47$ & $50$ & $7$ & $\mathbf{69}$ & $54$ & $60$ & $42$ & $55$ & $48$ \\ 
 & $35$ & $46$ & $50$ & $8$ & $\mathbf{69}$ & $54$ & $60$ & $43$ & $55$ & $48$ \\ 
\hline
\multirow{3}[2]{*}{$\beta(0.7,1)$} & $38$ & $\mathbf{49}$ & $13$ & $32$ & $9$ & $9$ & $14$ & $15$ & $28$ & $\mathbf{49}$ \\ 
 & $37$ & $48$ & $12$ & $32$ & $9$ & $9$ & $15$ & $16$ & $27$ & $\mathbf{50}$ \\ 
 & $37$ & $\mathbf{49}$ & $12$ & $31$ & $9$ & $9$ & $14$ & $15$ & $28$ & $\mathbf{49}$ \\ 
\hline
\multirow{3}[2]{*}{$\beta(1,1.5)$} & $60$ & $76$ & $63$ & $\mathbf{77}$ & $32$ & $48$ & $52$ & $61$ & $34$ & $70$ \\ 
 & $59$ & $75$ & $62$ & $\mathbf{77}$ & $33$ & $48$ & $52$ & $59$ & $33$ & $70$ \\ 
 & $59$ & $75$ & $63$ & $\mathbf{77}$ & $32$ & $48$ & $52$ & $60$ & $33$ & $70$ \\ 
\hline \\[-1.8ex] 
\end{tabular} 
\end{table} 

\begin{table}[!htbp] \centering 
  \caption{Estimated powers for $20\%$ censoring for a sample size of n=50 with three different censoring distributions.} 
  \label{table3} 
  \scriptsize
\begin{tabular}{@{\extracolsep{1pt}} ccccccccccc} 
\\[-1.8ex]\hline \vspace{-0.15cm}\\
$F$ & $\widetilde{KS}$ & $\widetilde{CM}$ & $\widetilde{CO}$ & $\widetilde{EP}$ & $\widetilde{L}_{.25}$ & $\widetilde{L}_{.5}$ & $\widetilde{B}_{.25}$ & $\widetilde{B}_{.5}$ & $\widetilde{H}_{.5}$ & $\widetilde{H}_{1}$ \\
\hline 
\multirow{3}[2]{*}{$Exp(1)$} & $4$ & $4$ & $5$ & $5$ & $5$ & $5$ & $4$ & $3$ & $4$ & $4$ \\ 
 & $4$ & $4$ & $5$ & $4$ & $5$ & $5$ & $4$ & $4$ & $4$ & $4$ \\ 
 & $4$ & $4$ & $5$ & $5$ & $5$ & $5$ & $4$ & $3$ & $4$ & $4$ \\ 
\hline
\multirow{3}[2]{*}{$\Gamma(0.6)$} & $47$ & $57$ & $79$ & $56$ & $\mathbf{80}$ & $77$ & $74$ & $67$ & $49$ & $49$ \\ 
& $48$ & $56$ & $\mathbf{80}$ & $55$ & $\mathbf{80}$ & $77$ & $73$ & $63$ & $48$ & $48$ \\ 
& $47$ & $56$ & $\mathbf{80}$ & $56$ & $79$ & $76$ & $73$ & $65$ & $49$ & $49$ \\ 
\hline
\multirow{3}[2]{*}{$\Gamma(0.8)$} & $11$ & $14$ & $23$ & $15$ & $\mathbf{26}$ & $23$ & $21$ & $17$ & $12$ & $13$ \\ 
& $11$ & $14$ & $23$ & $15$ & $\mathbf{26}$ & $23$ & $21$ & $17$ & $11$ & $12$ \\ 
& $11$ & $14$ & $24$ & $15$ & $\mathbf{26}$ & $23$ & $20$ & $16$ & $12$ & $13$ \\
\hline
\multirow{3}[2]{*}{$\Gamma(1.2)$} & $8$ & $10$ & $\mathbf{13}$ & $9$ & $11$ & $\mathbf{13}$ & $7$ & $5$ & $7$ & $9$ \\ 
 & $8$ & $10$ & $\mathbf{13}$ & $10$ & $11$ & $\mathbf{13}$ & $4$ & $2$ & $7$ & $9$ \\ 
 & $8$ & $9$ & $\mathbf{13}$ & $10$ & $11$ & $12$ & $5$ & $2$ & $7$ & $8$ \\ 
\hline
\multirow{3}[2]{*}{$W(0.8)$} & $25$ & $31$ & $\mathbf{46}$ & $35$ & $45$ & $44$ & $42$ & $37$ & $21$ & $26$ \\ 
 & $24$ & $30$ & $\mathbf{45}$ & $33$ & $44$ & $44$ & $42$ & $37$ & $21$ & $25$ \\ 
 & $24$ & $31$ & $\mathbf{46}$ & $34$ & $45$ & $44$ & $41$ & $35$ & $21$ & $26$ \\ 
\hline
\multirow{3}[2]{*}{$W(1.2)$} & $16$ & $20$ & $\mathbf{26}$ & $20$ & $21$ & $25$ & $14$ & $11$ & $12$ & $18$ \\ 
 & $16$ & $19$ & $\mathbf{26}$ & $20$ & $20$ & $24$ & $9$ & $5$ & $12$ & $18$ \\ 
 & $15$ & $20$ & $\mathbf{26}$ & $20$ & $20$ & $24$ & $11$ & $6$ & $12$ & $17$ \\ 
\hline
\multirow{3}[2]{*}{$LN(1)$} & $17$ & $21$ & $16$ & $11$ & $\mathbf{31}$ & $23$ & $20$ & $13$ & $24$ & $14$ \\ 
 & $14$ & $18$ & $16$ & $8$ & $\mathbf{34}$ & $25$ & $16$ & $9$ & $23$ & $13$ \\ 
 & $15$ & $18$ & $16$ & $9$ & $\mathbf{32}$ & $24$ & $16$ & $10$ & $22$ & $13$ \\ 
\hline
\multirow{3}[2]{*}{$LN(1.5)$} & $66$ & $73$ & $67$ & $\mathbf{78}$ & $53$ & $66$ & $72$ & $74$ & $40$ & $62$ \\ 
 & $60$ & $69$ & $60$ & $\mathbf{72}$ & $45$ & $59$ & $69$ & $69$ & $33$ & $57$ \\ 
 & $60$ & $69$ & $62$ & $\mathbf{72}$ & $46$ & $60$ & $67$ & $67$ & $34$ & $57$ \\ 
\hline
\multirow{3}[2]{*}{$\chi^2(1)$} & $75$ & $82$ & $\mathbf{96}$ & $79$ & $\mathbf{96}$ & $94$ & $94$ & $90$ & $76$ & $75$ \\ 
 & $75$ & $82$ & $\mathbf{96}$ & $79$ & $\mathbf{96}$ & $94$ & $93$ & $88$ & $76$ & $75$ \\ 
 & $75$ & $82$ & $\mathbf{96}$ & $79$ & $\mathbf{96}$ & $94$ & $93$ & $88$ & $76$ & $75$ \\ 
\hline
\multirow{3}[2]{*}{$\chi^2(3)$} & $26$ & $34$ & $\mathbf{48}$ & $32$ & $43$ & $46$ & $31$ & $23$ & $24$ & $31$ \\ 
 & $26$ & $32$ & $\mathbf{47}$ & $32$ & $42$ & $46$ & $21$ & $10$ & $22$ & $30$ \\ 
 & $25$ & $32$ & $46$ & $30$ & $43$ & $\mathbf{47}$ & $19$ & $9$ & $22$ & $29$ \\ 
\hline
\multirow{3}[2]{*}{$\beta(1,1)$} & $85$ & $93$ & $77$ & $84$ & $48$ & $68$ & $72$ & $80$ & $70$ & $\mathbf{94}$ \\ 
 & $85$ & $93$ & $77$ & $80$ & $48$ & $66$ & $68$ & $76$ & $69$ & $\mathbf{94}$ \\ 
 & $85$ & $93$ & $77$ & $82$ & $48$ & $67$ & $70$ & $76$ & $70$ & $\mathbf{94}$ \\ 
\hline
\multirow{3}[2]{*}{$\beta(0.5,1)$} & $33$ & $44$ & $56$ & $9$ & $\mathbf{70}$ & $57$ & $58$ & $39$ & $49$ & $42$ \\ 
 & $31$ & $43$ & $58$ & $10$ & $\mathbf{71}$ & $58$ & $55$ & $33$ & $49$ & $41$ \\ 
 & $33$ & $43$ & $57$ & $9$ & $\mathbf{71}$ & $57$ & $56$ & $38$ & $49$ & $42$ \\ 
\hline
\multirow{3}[2]{*}{$\beta(0.7,1)$} & $30$ & $38$ & $9$ & $21$ & $9$ & $8$ & $9$ & $8$ & $23$ & $\mathbf{40}$ \\ 
 & $29$ & $35$ & $9$ & $18$ & $10$ & $8$ & $8$ & $6$ & $21$ & $\mathbf{39}$ \\ 
 & $29$ & $37$ & $9$ & $19$ & $10$ & $8$ & $8$ & $8$ & $22$ & $\mathbf{39}$ \\ 
\hline
\multirow{3}[2]{*}{$\beta(1,1.5)$} & $49$ & $\mathbf{65}$ & $52$ & $59$ & $27$ & $41$ & $37$ & $43$ & $27$ & $60$ \\ 
 & $48$ & $\mathbf{62}$ & $51$ & $57$ & $28$ & $40$ & $32$ & $33$ & $26$ & $58$ \\ 
 & $49$ & $\mathbf{63}$ & $52$ & $59$ & $28$ & $41$ & $36$ & $39$ & $26$ & $60$ \\ 
\hline \\[-1.8ex] 
\end{tabular} 
\end{table}

\begin{table}[!htbp] \centering 
  \caption{Estimated powers for $30\%$ censoring for a sample size of n=50 with three different censoring distributions.} 
  \label{table4} 
  \scriptsize
\begin{tabular}{@{\extracolsep{1pt}} ccccccccccc} 
\\[-1.8ex]\hline \vspace{-0.15cm}\\
$F$ & $\widetilde{KS}$ & $\widetilde{CM}$ & $\widetilde{CO}$ & $\widetilde{EP}$ & $\widetilde{L}_{.25}$ & $\widetilde{L}_{.5}$ & $\widetilde{B}_{.25}$ & $\widetilde{B}_{.5}$ & $\widetilde{H}_{.5}$ & $\widetilde{H}_{1}$ \\
\hline 
\multirow{3}[2]{*}{$Exp(1)$} & $3$ & $3$ & $5$ & $3$ & $5$ & $5$ & $3$ & $3$ & $4$ & $4$ \\ 
 & $3$ & $3$ & $5$ & $2$ & $5$ & $5$ & $4$ & $4$ & $4$ & $4$ \\ 
 & $3$ & $3$ & $5$ & $3$ & $5$ & $5$ & $3$ & $3$ & $4$ & $4$ \\ 
\hline
\multirow{3}[2]{*}{$\Gamma(0.6)$} & $33$ & $46$ & $76$ & $42$ & $\mathbf{77}$ & $73$ & $62$ & $49$ & $38$ & $40$ \\ 
& $33$ & $42$ & $\mathbf{77}$ & $33$ & $76$ & $73$ & $58$ & $45$ & $35$ & $38$ \\ 
& $33$ & $45$ & $\mathbf{77}$ & $41$ & $76$ & $72$ & $60$ & $46$ & $38$ & $40$ \\ 
\hline
\multirow{3}[2]{*}{$\Gamma(0.8)$} & $7$ & $10$ & $22$ & $11$ & $\mathbf{24}$ & $21$ & $15$ & $11$ & $9$ & $10$ \\ 
& $7$ & $9$ & $22$ & $6$ & $\mathbf{25}$ & $21$ & $16$ & $13$ & $8$ & $9$ \\ 
& $7$ & $10$ & $23$ & $10$ & $\mathbf{25}$ & $21$ & $14$ & $10$ & $9$ & $10$ \\ 
\hline
\multirow{3}[2]{*}{$\Gamma(1.2)$} & $5$ & $7$ & $\mathbf{12}$ & $6$ & $10$ & $11$ & $2$ & $1$ & $5$ & $7$ \\ 
 & $5$ & $4$ & $\mathbf{12}$ & $5$ & $10$ & $11$ & $1$ & $1$ & $5$ & $6$ \\ 
 & $5$ & $6$ & $\mathbf{12}$ & $5$ & $10$ & $11$ & $1$ & $1$ & $5$ & $6$ \\ 
\hline
\multirow{3}[2]{*}{$W(0.8)$} & $16$ & $23$ & $\mathbf{41}$ & $25$ & $\mathbf{41}$ & $40$ & $32$ & $24$ & $16$ & $20$ \\ 
 & $14$ & $20$ & $\mathbf{40}$ & $15$ & $\mathbf{40}$ & $38$ & $30$ & $25$ & $15$ & $18$ \\ 
 & $15$ & $22$ & $\mathbf{41}$ & $22$ & $\mathbf{41}$ & $39$ & $29$ & $22$ & $16$ & $20$ \\ 
\hline
\multirow{3}[2]{*}{$W(1.2)$} & $9$ & $13$ & $\mathbf{23}$ & $11$ & $19$ & $21$ & $5$ & $1$ & $8$ & $12$ \\ 
 & $9$ & $9$ & $\mathbf{23}$ & $9$ & $18$ & $20$ & $1$ & $0$ & $8$ & $12$ \\ 
 & $9$ & $11$ & $\mathbf{23}$ & $10$ & $18$ & $21$ & $2$ & $1$ & $8$ & $12$ \\ 
\hline
\multirow{3}[2]{*}{$LN(1)$} & $10$ & $14$ & $21$ & $8$ & $\mathbf{34}$ & $26$ & $10$ & $6$ & $16$ & $10$ \\ 
 & $7$ & $9$ & $23$ & $6$ & $\mathbf{39}$ & $29$ & $4$ & $3$ & $16$ & $9$ \\ 
 & $8$ & $11$ & $21$ & $7$ & $\mathbf{36}$ & $27$ & $6$ & $4$ & $15$ & $9$ \\ 
\hline
\multirow{3}[2]{*}{$LN(1.5)$} & $41$ & $53$ & $47$ & $\mathbf{55}$ & $36$ & $48$ & $52$ & $51$ & $22$ & $42$ \\ 
 & $32$ & $42$ & $31$ & $29$ & $25$ & $37$ & $\mathbf{46}$ & $44$ & $15$ & $34$ \\ 
 & $35$ & $\mathbf{46}$ & $40$ & $41$ & $29$ & $42$ & $\mathbf{46}$ & $44$ & $18$ & $37$ \\ 
\hline
\multirow{3}[2]{*}{$\chi^2(1)$} & $59$ & $74$ & $\mathbf{95}$ & $65$ & $94$ & $92$ & $87$ & $76$ & $65$ & $65$ \\ 
 & $61$ & $71$ & $\mathbf{94}$ & $55$ & $\mathbf{94}$ & $92$ & $83$ & $70$ & $63$ & $64$ \\ 
 & $59$ & $74$ & $\mathbf{95}$ & $63$ & $94$ & $92$ & $86$ & $73$ & $64$ & $65$ \\ 
\hline
\multirow{3}[2]{*}{$\chi^2(3)$} & $16$ & $23$ & $\mathbf{44}$ & $18$ & $40$ & $42$ & $12$ & $3$ & $16$ & $21$ \\ 
 & $15$ & $15$ & $\mathbf{42}$ & $16$ & $39$ & $39$ & $1$ & $0$ & $15$ & $21$ \\ 
 & $13$ & $17$ & $\mathbf{43}$ & $15$ & $39$ & $41$ & $3$ & $1$ & $13$ & $19$ \\ 
\hline
\multirow{3}[2]{*}{$\beta(1,1)$} & $76$ & $85$ & $67$ & $54$ & $42$ & $59$ & $55$ & $59$ & $58$ & $\mathbf{87}$ \\ 
 & $75$ & $79$ & $64$ & $39$ & $40$ & $55$ & $37$ & $31$ & $51$ & $\mathbf{84}$ \\ 
 & $77$ & $84$ & $66$ & $49$ & $41$ & $57$ & $51$ & $53$ & $56$ & $\mathbf{86}$ \\ 
\hline
\multirow{3}[2]{*}{$\beta(0.5,1)$} & $25$ & $37$ & $63$ & $9$ & $\mathbf{72}$ & $59$ & $47$ & $25$ & $40$ & $34$ \\ 
 & $21$ & $29$ & $67$ & $12$ & $\mathbf{74}$ & $63$ & $36$ & $14$ & $39$ & $30$ \\ 
 & $24$ & $35$ & $64$ & $10$ & $\mathbf{73}$ & $59$ & $46$ & $23$ & $40$ & $33$ \\ 
\hline
\multirow{3}[2]{*}{$\beta(0.7,1)$} & $21$ & $24$ & $7$ & $9$ & $10$ & $7$ & $4$ & $2$ & $16$ & $\mathbf{27}$ \\ 
 & $18$ & $15$ & $8$ & $5$ & $11$ & $7$ & $1$ & $0$ & $12$ & $\mathbf{21}$ \\ 
 & $20$ & $22$ & $8$ & $8$ & $10$ & $7$ & $4$ & $1$ & $15$ & $\mathbf{26}$ \\ 
\hline
\multirow{3}[2]{*}{$\beta(1,1.5)$} & $35$ & $\mathbf{46}$ & $42$ & $33$ & $23$ & $33$ & $18$ & $14$ & $19$ & $45$ \\ 
 & $31$ & $34$ & $\mathbf{38}$ & $21$ & $22$ & $30$ & $5$ & $1$ & $14$ & $\mathbf{38}$ \\ 
 & $34$ & $\mathbf{45}$ & $42$ & $30$ & $23$ & $33$ & $15$ & $10$ & $17$ & $43$ \\ 
\hline \\[-1.8ex] 
\end{tabular} 
\end{table}

\begin{table}[!htbp] \centering 
  \caption{Estimated powers for $10\%$ censoring for a sample size of n=100 with three different censoring distributions.} 
  \label{table5} 
  \scriptsize
\begin{tabular}{@{\extracolsep{1pt}} ccccccccccc} 
\\[-1.8ex]\hline \vspace{-0.15cm}\\
$F$ & $\widetilde{KS}$ & $\widetilde{CM}$ & $\widetilde{CO}$ & $\widetilde{EP}$ & $\widetilde{L}_{.25}$ & $\widetilde{L}_{.5}$ & $\widetilde{B}_{.25}$ & $\widetilde{B}_{.5}$ & $\widetilde{H}_{.5}$ & $\widetilde{H}_{1}$ \\
\hline 
\multirow{3}[2]{*}{$Exp(1)$} & $5$ & $5$ & $5$ & $5$ & $5$ & $5$ & $5$ & $4$ & $5$ & $5$ \\ 
 & $5$ & $5$ & $5$ & $5$ & $5$ & $5$ & $5$ & $4$ & $5$ & $5$ \\ 
 & $5$ & $5$ & $5$ & $5$ & $5$ & $5$ & $4$ & $4$ & $5$ & $5$ \\ 
\hline
\multirow{3}[2]{*}{$\Gamma(0.6)$} & $85$ & $90$ & $\mathbf{98}$ & $89$ & $97$ & $97$ & $97$ & $96$ & $85$ & $86$ \\ 
& $85$ & $90$ & $\mathbf{98}$ & $89$ & $\mathbf{98}$ & $97$ & $97$ & $96$ & $85$ & $86$ \\ 
& $85$ & $90$ & $\mathbf{97}$ & $89$ & $\mathbf{97}$ & $\mathbf{97}$ & $\mathbf{97}$ & $96$ & $86$ & $86$ \\ 
\hline
\multirow{3}[2]{*}{$\Gamma(0.8)$} & $24$ & $28$ & $42$ & $31$ & $\mathbf{44}$ & $41$ & $41$ & $37$ & $24$ & $25$ \\ 
& $24$ & $28$ & $43$ & $30$ & $\mathbf{45}$ & $41$ & $41$ & $37$ & $24$ & $25$ \\ 
& $24$ & $29$ & $43$ & $30$ & $\mathbf{44}$ & $42$ & $41$ & $38$ & $24$ & $26$ \\
\hline
\multirow{3}[2]{*}{$\Gamma(1.2)$} & $16$ & $19$ & $\mathbf{26}$ & $20$ & $23$ & $25$ & $22$ & $21$ & $15$ & $18$ \\ 
 & $16$ & $19$ & $\mathbf{25}$ & $20$ & $22$ & $24$ & $21$ & $19$ & $15$ & $18$ \\ 
 & $16$ & $20$ & $\mathbf{25}$ & $20$ & $22$ & $24$ & $20$ & $18$ & $15$ & $18$ \\ 
\hline
\multirow{3}[2]{*}{$W(0.8)$} & $56$ & $65$ & $\mathbf{78}$ & $69$ & $75$ & $75$ & $76$ & $75$ & $49$ & $57$ \\ 
 & $56$ & $64$ & $\mathbf{78}$ & $69$ & $74$ & $75$ & $76$ & $74$ & $49$ & $57$ \\ 
 & $56$ & $64$ & $\mathbf{77}$ & $69$ & $74$ & $75$ & $75$ & $74$ & $50$ & $57$ \\ 
\hline
\multirow{3}[2]{*}{$W(1.2)$} & $38$ & $47$ & $\mathbf{56}$ & $51$ & $46$ & $51$ & $48$ & $49$ & $29$ & $42$ \\ 
 & $39$ & $46$ & $\mathbf{56}$ & $51$ & $46$ & $52$ & $48$ & $47$ & $29$ & $41$ \\ 
 & $38$ & $47$ & $\mathbf{55}$ & $51$ & $47$ & $52$ & $46$ & $45$ & $30$ & $41$ \\ 
\hline
\multirow{3}[2]{*}{$LN(1)$} & $40$ & $50$ & $18$ & $14$ & $\mathbf{58}$ & $36$ & $\mathbf{58}$ & $36$ & $\mathbf{58}$ & $30$ \\ 
 & $38$ & $48$ & $16$ & $11$ & $\mathbf{60}$ & $37$ & $57$ & $37$ & $58$ & $31$ \\ 
 & $38$ & $48$ & $18$ & $11$ & $\mathbf{61}$ & $39$ & $55$ & $34$ & $58$ & $30$ \\ 
\hline
\multirow{3}[2]{*}{$LN(1.5)$} & $98$ & $\mathbf{99}$ & $98$ & $\mathbf{99}$ & $93$ & $97$ & $98$ & $\mathbf{99}$ & $88$ & $97$ \\ 
 & $98$ & $\mathbf{99}$ & $98$ & $\mathbf{99}$ & $92$ & $97$ & $98$ & $\mathbf{99}$ & $87$ & $97$ \\ 
 & $98$ & $\mathbf{99}$ & $98$ & $\mathbf{99}$ & $91$ & $97$ & $98$ & $\mathbf{99}$ & $86$ & $96$ \\ 
\hline
\multirow{3}[2]{*}{$\chi^2(1)$} & $98$ & $99$ & $\mathbf{100}$ & $99$ & $\mathbf{100}$ & $\mathbf{100}$ & $\mathbf{100}$ & $\mathbf{100}$ & $98$ & $98$ \\ 
 & $99$ & $99$ & $\mathbf{100}$ & $99$ & $\mathbf{100}$ & $\mathbf{100}$ & $\mathbf{100}$ & $\mathbf{100}$ & $98$ & $98$ \\ 
 & $99$ & $99$ & $\mathbf{100}$ & $99$ & $\mathbf{100}$ & $\mathbf{100}$ & $\mathbf{100}$ & $\mathbf{100}$ & $99$ & $98$ \\ 
\hline
\multirow{3}[2]{*}{$\chi^2(3)$} & $62$ & $71$ & $\mathbf{84}$ & $71$ & $81$ & $83$ & $80$ & $78$ & $58$ & $67$ \\ 
 & $62$ & $71$ & $\mathbf{83}$ & $71$ & $81$ & $\mathbf{83}$ & $79$ & $75$ & $58$ & $66$ \\ 
 & $62$ & $71$ & $\mathbf{84}$ & $71$ & $80$ & $82$ & $75$ & $71$ & $57$ & $65$ \\ 
\hline
\multirow{3}[2]{*}{$\beta(1,1)$} & $\mathbf{100}$ & $\mathbf{100}$ & $99$ & $\mathbf{100}$ & $87$ & $97$ & $99$ & $\mathbf{100}$ & $99$ & $\mathbf{100}$ \\ 
 & $\mathbf{100}$ & $\mathbf{100}$ & $99$ & $\mathbf{100}$ & $86$ & $97$ & $99$ & $\mathbf{100}$ & $99$ & $\mathbf{100}$ \\ 
 & $\mathbf{100}$ & $\mathbf{100}$ & $99$ & $\mathbf{100}$ & $87$ & $97$ & $99$ & $\mathbf{100}$ & $99$ & $\mathbf{100}$ \\ 
\hline
\multirow{3}[2]{*}{$\beta(0.5,1)$} & $68$ & $83$ & $73$ & $9$ & $\mathbf{91}$ & $80$ & $87$ & $74$ & $87$ & $82$ \\ 
 & $67$ & $83$ & $73$ & $9$ & $\mathbf{91}$ & $79$ & $87$ & $73$ & $86$ & $82$ \\ 
 & $67$ & $83$ & $73$ & $9$ & $\mathbf{91}$ & $80$ & $87$ & $73$ & $86$ & $82$ \\ 
\hline
\multirow{3}[2]{*}{$\beta(0.7,1)$} & $74$ & $87$ & $18$ & $62$ & $10$ & $12$ & $32$ & $34$ & $60$ & $\mathbf{88}$ \\ 
 & $74$ & $87$ & $18$ & $61$ & $11$ & $12$ & $31$ & $34$ & $59$ & $\mathbf{88}$ \\ 
 & $74$ & $86$ & $18$ & $60$ & $11$ & $12$ & $32$ & $32$ & $58$ & $\mathbf{88}$ \\ 
\hline
\multirow{3}[2]{*}{$\beta(1,1.5)$} & $91$ & $\mathbf{98}$ & $91$ & $\mathbf{98}$ & $61$ & $80$ & $88$ & $92$ & $68$ & $97$ \\ 
 & $92$ & $\mathbf{98}$ & $91$ & $\mathbf{98}$ & $61$ & $81$ & $88$ & $92$ & $68$ & $97$ \\ 
 & $91$ & $\mathbf{98}$ & $91$ & $\mathbf{98}$ & $61$ & $81$ & $87$ & $92$ & $69$ & $97$ \\ 
\hline \\[-1.8ex] 
\end{tabular} 
\end{table}

\begin{table}[!htbp] \centering 
  \caption{Estimated powers for $20\%$ censoring for a sample size of n=100 with three different censoring distributions.} 
  \label{table6} 
  \scriptsize
\begin{tabular}{@{\extracolsep{1pt}} ccccccccccc} 
\\[-1.8ex]\hline \vspace{-0.15cm}\\
$F$ & $\widetilde{KS}$ & $\widetilde{CM}$ & $\widetilde{CO}$ & $\widetilde{EP}$ & $\widetilde{L}_{.25}$ & $\widetilde{L}_{.5}$ & $\widetilde{B}_{.25}$ & $\widetilde{B}_{.5}$ & $\widetilde{H}_{.5}$ & $\widetilde{H}_{1}$ \\
\hline 
\multirow{3}[2]{*}{$Exp(1)$} & $5$ & $5$ & $5$ & $5$ & $5$ & $5$ & $4$ & $3$ & $5$ & $5$ \\ 
 & $4$ & $5$ & $5$ & $5$ & $5$ & $5$ & $4$ & $4$ & $4$ & $4$ \\ 
 & $4$ & $5$ & $5$ & $5$ & $5$ & $5$ & $3$ & $3$ & $5$ & $5$ \\ 
\hline
\multirow{3}[2]{*}{$\Gamma(0.6)$} & $80$ & $86$ & $\mathbf{97}$ & $84$ & $\mathbf{97}$ & $96$ & $95$ & $92$ & $81$ & $80$ \\ 
& $80$ & $86$ & $\mathbf{97}$ & $85$ & $\mathbf{97}$ & $96$ & $94$ & $90$ & $80$ & $79$ \\ 
& $79$ & $86$ & $\mathbf{97}$ & $84$ & $96$ & $96$ & $95$ & $92$ & $81$ & $80$ \\ 
\hline
\multirow{3}[2]{*}{$\Gamma(0.8)$} & $20$ & $25$ & $41$ & $27$ & $\mathbf{42}$ & $39$ & $36$ & $30$ & $21$ & $22$ \\ 
& $20$ & $25$ & $41$ & $26$ & $\mathbf{42}$ & $38$ & $34$ & $28$ & $20$ & $22$ \\ 
& $20$ & $26$ & $41$ & $27$ & $\mathbf{42}$ & $39$ & $34$ & $29$ & $21$ & $22$ \\ 
\hline
\multirow{3}[2]{*}{$\Gamma(1.2)$} & $14$ & $17$ & $\mathbf{24}$ & $18$ & $22$ & $23$ & $15$ & $11$ & $13$ & $15$ \\ 
 & $13$ & $16$ & $\mathbf{24}$ & $18$ & $21$ & $23$ & $9$ & $4$ & $12$ & $14$ \\ 
 & $13$ & $17$ & $\mathbf{23}$ & $17$ & $21$ & $22$ & $12$ & $6$ & $12$ & $14$ \\ 
\hline
\multirow{3}[2]{*}{$W(0.8)$} & $47$ & $57$ & $\mathbf{73}$ & $61$ & $70$ & $71$ & $69$ & $65$ & $43$ & $49$ \\ 
 & $46$ & $55$ & $\mathbf{73}$ & $59$ & $69$ & $70$ & $67$ & $61$ & $42$ & $48$ \\ 
 & $46$ & $56$ & $\mathbf{72}$ & $60$ & $70$ & $69$ & $66$ & $61$ & $42$ & $50$ \\ 
\hline
\multirow{3}[2]{*}{$W(1.2)$} & $31$ & $40$ & $\mathbf{51}$ & $42$ & $42$ & $47$ & $35$ & $29$ & $24$ & $34$ \\ 
 & $31$ & $39$ & $\mathbf{50}$ & $42$ & $42$ & $47$ & $22$ & $13$ & $23$ & $33$ \\ 
 & $31$ & $39$ & $\mathbf{50}$ & $43$ & $43$ & $47$ & $30$ & $20$ & $24$ & $34$ \\ 
\hline
\multirow{3}[2]{*}{$LN(1)$} & $35$ & $45$ & $25$ & $12$ & $\mathbf{65}$ & $44$ & $53$ & $30$ & $53$ & $27$ \\ 
 & $34$ & $41$ & $28$ & $8$ & $\mathbf{70}$ & $49$ & $46$ & $24$ & $53$ & $27$ \\ 
 & $33$ & $43$ & $29$ & $10$ & $\mathbf{69}$ & $48$ & $46$ & $23$ & $52$ & $26$ \\ 
\hline
\multirow{3}[2]{*}{$LN(1.5)$} & $92$ & $95$ & $92$ & $\mathbf{97}$ & $81$ & $91$ & $95$ & $95$ & $72$ & $91$ \\ 
 & $90$ & $94$ & $89$ & $\mathbf{96}$ & $74$ & $87$ & $92$ & $93$ & $65$ & $88$ \\ 
 & $90$ & $94$ & $88$ & $\mathbf{95}$ & $74$ & $87$ & $92$ & $92$ & $67$ & $88$ \\ 
\hline
\multirow{3}[2]{*}{$\chi^2(1)$} & $97$ & $99$ & $\mathbf{100}$ & $98$ & $\mathbf{100}$ & $\mathbf{100}$ & $\mathbf{100}$ & $\mathbf{100}$ & $97$ & $97$ \\ 
 & $97$ & $99$ & $\mathbf{100}$ & $98$ & $\mathbf{100}$ & $\mathbf{100}$ & $\mathbf{100}$ & $99$ & $97$ & $97$ \\ 
 & $97$ & $99$ & $\mathbf{100}$ & $98$ & $\mathbf{100}$ & $\mathbf{100}$ & $\mathbf{100}$ & $99$ & $98$ & $97$ \\ 
\hline
\multirow{3}[2]{*}{$\chi^2(3)$} & $54$ & $65$ & $\mathbf{81}$ & $63$ & $78$ & $80$ & $68$ & $58$ & $51$ & $58$ \\ 
 & $52$ & $64$ & $\mathbf{80}$ & $63$ & $78$ & $79$ & $51$ & $32$ & $49$ & $56$ \\ 
 & $50$ & $63$ & $\mathbf{81}$ & $64$ & $77$ & $79$ & $51$ & $32$ & $47$ & $54$ \\ 
\hline
\multirow{3}[2]{*}{$\beta(1,1)$} & $\mathbf{100}$ & $\mathbf{100}$ & $97$ & $99$ & $82$ & $94$ & $98$ & $99$ & $98$ & $\mathbf{100}$ \\ 
 & $\mathbf{100}$ & $\mathbf{100}$ & $97$ & $99$ & $80$ & $93$ & $97$ & $99$ & $98$ & $\mathbf{100}$ \\ 
 & $\mathbf{100}$ & $\mathbf{100}$ & $97$ & $99$ & $81$ & $94$ & $98$ & $99$ & $98$ & $\mathbf{100}$ \\ 
\hline
\multirow{3}[2]{*}{$\beta(0.5,1)$} & $66$ & $80$ & $80$ & $11$ & $\mathbf{92}$ & $83$ & $86$ & $71$ & $83$ & $76$ \\ 
 & $65$ & $79$ & $82$ & $15$ & $\mathbf{92}$ & $83$ & $84$ & $66$ & $82$ & $75$ \\ 
 & $66$ & $80$ & $81$ & $13$ & $\mathbf{92}$ & $83$ & $86$ & $69$ & $82$ & $76$ \\ 
\hline
\multirow{3}[2]{*}{$\beta(0.7,1)$} & $69$ & $78$ & $11$ & $45$ & $11$ & $9$ & $20$ & $21$ & $50$ & $\mathbf{80}$ \\ 
 & $68$ & $75$ & $11$ & $39$ & $12$ & $9$ & $16$ & $15$ & $49$ & $\mathbf{78}$ \\ 
 & $69$ & $76$ & $11$ & $42$ & $11$ & $9$ & $20$ & $18$ & $51$ & $\mathbf{79}$ \\ 
\hline
\multirow{3}[2]{*}{$\beta(1,1.5)$} & $86$ & $\mathbf{95}$ & $84$ & $94$ & $54$ & $73$ & $77$ & $83$ & $60$ & $94$ \\ 
 & $86$ & $\mathbf{94}$ & $82$ & $92$ & $53$ & $72$ & $71$ & $76$ & $58$ & $93$ \\ 
 & $86$ & $\mathbf{95}$ & $83$ & $93$ & $54$ & $73$ & $74$ & $81$ & $59$ & $93$ \\ 
\hline \\[-1.8ex] 
\end{tabular} 
\end{table}

\begin{table}[!htbp] \centering 
  \caption{Estimated powers for $30\%$ censoring for a sample size of n=100 with three different censoring distributions.} 
  \label{table7} 
  \scriptsize
\begin{tabular}{@{\extracolsep{1pt}} ccccccccccc} 
\\[-1.8ex]\hline \vspace{-0.15cm}\\
$F$ & $\widetilde{KS}$ & $\widetilde{CM}$ & $\widetilde{CO}$ & $\widetilde{EP}$ & $\widetilde{L}_{.25}$ & $\widetilde{L}_{.5}$ & $\widetilde{B}_{.25}$ & $\widetilde{B}_{.5}$ & $\widetilde{H}_{.5}$ & $\widetilde{H}_{1}$ \\
\hline 
\multirow{3}[2]{*}{$Exp(1)$} & $4$ & $4$ & $5$ & $4$ & $5$ & $5$ & $2$ & $2$ & $4$ & $4$ \\ 
 & $3$ & $3$ & $5$ & $3$ & $5$ & $5$ & $4$ & $3$ & $4$ & $4$ \\ 
 & $3$ & $4$ & $5$ & $4$ & $5$ & $5$ & $3$ & $2$ & $4$ & $4$ \\ 
\hline
\multirow{3}[2]{*}{$\Gamma(0.6)$} & $66$ & $80$ & $\mathbf{95}$ & $73$ & $\mathbf{95}$ & $94$ & $90$ & $80$ & $71$ & $71$ \\ 
& $66$ & $78$ & $\mathbf{96}$ & $65$ & $95$ & $94$ & $82$ & $67$ & $68$ & $69$ \\ 
& $66$ & $80$ & $\mathbf{96}$ & $74$ & $95$ & $94$ & $88$ & $77$ & $71$ & $70$ \\ 
\hline
\multirow{3}[2]{*}{$\Gamma(0.8)$}& $14$ & $20$ & $39$ & $22$ & $\mathbf{40}$ & $37$ & $26$ & $17$ & $17$ & $18$ \\ 
& $12$ & $19$ & $\mathbf{39}$ & $14$ & $\mathbf{39}$ & $36$ & $22$ & $18$ & $14$ & $15$ \\ 
& $14$ & $20$ & $\mathbf{39}$ & $21$ & $\mathbf{39}$ & $36$ & $24$ & $15$ & $16$ & $18$ \\
\hline
\multirow{3}[2]{*}{$\Gamma(1.2)$} & $9$ & $13$ & $\mathbf{23}$ & $13$ & $21$ & $21$ & $6$ & $2$ & $9$ & $11$ \\ 
 & $7$ & $8$ & $\mathbf{22}$ & $13$ & $19$ & $20$ & $1$ & $1$ & $8$ & $10$ \\ 
 & $7$ & $12$ & $\mathbf{21}$ & $12$ & $20$ & $20$ & $3$ & $1$ & $8$ & $10$ \\ 
\hline
\multirow{3}[2]{*}{$W(0.8)$} & $34$ & $46$ & $\mathbf{67}$ & $48$ & $65$ & $65$ & $55$ & $44$ & $33$ & $39$ \\ 
 & $30$ & $42$ & $\mathbf{65}$ & $35$ & $63$ & $63$ & $47$ & $37$ & $28$ & $35$ \\ 
 & $31$ & $45$ & $\mathbf{66}$ & $46$ & $65$ & $64$ & $50$ & $38$ & $32$ & $38$ \\ 
\hline
\multirow{3}[2]{*}{$W(1.2)$} & $20$ & $31$ & $\mathbf{46}$ & $30$ & $38$ & $42$ & $15$ & $5$ & $16$ & $25$ \\ 
 & $17$ & $21$ & $\mathbf{44}$ & $28$ & $37$ & $39$ & $1$ & $0$ & $14$ & $23$ \\ 
 & $18$ & $29$ & $\mathbf{45}$ & $29$ & $39$ & $42$ & $8$ & $2$ & $16$ & $23$ \\ 
\hline
\multirow{3}[2]{*}{$LN(1)$} & $24$ & $37$ & $35$ & $10$ & $\mathbf{71}$ & $51$ & $34$ & $15$ & $41$ & $21$ \\ 
 & $18$ & $28$ & $44$ & $14$ & $\mathbf{77}$ & $60$ & $11$ & $5$ & $38$ & $19$ \\ 
 & $19$ & $33$ & $40$ & $11$ & $\mathbf{74}$ & $56$ & $23$ & $8$ & $38$ & $20$ \\ 
\hline
\multirow{3}[2]{*}{$LN(1.5)$} & $75$ & $85$ & $76$ & $\mathbf{87}$ & $61$ & $77$ & $81$ & $80$ & $49$ & $75$ \\ 
 & $65$ & $\mathbf{77}$ & $59$ & $62$ & $46$ & $64$ & $70$ & $66$ & $35$ & $66$ \\ 
 & $69$ & $\mathbf{81}$ & $67$ & $77$ & $52$ & $70$ & $71$ & $67$ & $41$ & $69$ \\ 
\hline
\multirow{3}[2]{*}{$\chi^2(1)$} & $93$ & $97$ & $\mathbf{100}$ & $93$ & $\mathbf{100}$ & $\mathbf{100}$ & $99$ & $98$ & $94$ & $93$ \\ 
 & $93$ & $97$ & $\mathbf{100}$ & $88$ & $\mathbf{100}$ & $\mathbf{100}$ & $98$ & $92$ & $93$ & $92$ \\ 
 & $92$ & $97$ & $\mathbf{100}$ & $93$ & $\mathbf{100}$ & $\mathbf{100}$ & $99$ & $96$ & $94$ & $93$ \\ 
\hline
\multirow{3}[2]{*}{$\chi^2(3)$} & $38$ & $54$ & $\mathbf{77}$ & $47$ & $75$ & $76$ & $39$ & $16$ & $38$ & $45$ \\ 
 & $31$ & $41$ & $\mathbf{75}$ & $46$ & $73$ & $73$ & $4$ & $0$ & $34$ & $41$ \\ 
 & $28$ & $46$ & $\mathbf{76}$ & $44$ & $73$ & $74$ & $14$ & $2$ & $31$ & $39$ \\ 
\hline
\multirow{3}[2]{*}{$\beta(1,1)$} & $\mathbf{100}$ & $\mathbf{100}$ & $93$ & $90$ & $74$ & $90$ & $95$ & $97$ & $96$ & $\mathbf{100}$ \\ 
 & $\mathbf{100}$ & $99$ & $92$ & $78$ & $72$ & $86$ & $88$ & $92$ & $94$ & $\mathbf{100}$ \\ 
 & $\mathbf{100}$ & $\mathbf{100}$ & $93$ & $87$ & $73$ & $88$ & $93$ & $97$ & $95$ & $\mathbf{100}$ \\ 
\hline
\multirow{3}[2]{*}{$\beta(0.5,1)$} & $56$ & $75$ & $86$ & $14$ & $\mathbf{93}$ & $85$ & $81$ & $57$ & $75$ & $66$ \\ 
 & $48$ & $66$ & $90$ & $28$ & $\mathbf{94}$ & $87$ & $59$ & $22$ & $71$ & $58$ \\ 
 & $56$ & $75$ & $87$ & $16$ & $\mathbf{93}$ & $85$ & $79$ & $52$ & $75$ & $65$ \\ 
\hline
\multirow{3}[2]{*}{$\beta(0.7,1)$} & $59$ & $62$ & $8$ & $25$ & $12$ & $8$ & $9$ & $8$ & $37$ & $\mathbf{64}$ \\ 
 & $\mathbf{52}$ & $46$ & $8$ & $12$ & $13$ & $8$ & $2$ & $0$ & $29$ & $\mathbf{52}$ \\ 
 & $58$ & $60$ & $8$ & $22$ & $12$ & $8$ & $8$ & $5$ & $36$ & $\mathbf{63}$ \\ 
\hline
\multirow{3}[2]{*}{$\beta(1,1.5)$} & $78$ & $\mathbf{86}$ & $71$ & $74$ & $46$ & $63$ & $54$ & $55$ & $45$ & $85$ \\ 
 & $73$ & $76$ & $67$ & $54$ & $42$ & $58$ & $12$ & $3$ & $36$ & $\mathbf{77}$ \\ 
 & $78$ & $\mathbf{86}$ & $71$ & $71$ & $45$ & $62$ & $46$ & $46$ & $44$ & $84$ \\ 
\hline \\[-1.8ex] 
\end{tabular} 
\end{table} 

\end{document}